\documentclass[a4paper,11pt]{article}
\pdfoutput=1
\usepackage{jheppub}
\usepackage{xcolor}
\usepackage{amsmath}
\usepackage[T1]{fontenc}
\usepackage{amssymb}

\usepackage{float}
\usepackage{comment}

\def\lr#1{\left(#1\right)}
\def\slr#1{\left[#1\right]}

\def\RS2{\mathbb R\times S^2_F}

\def \be  {\begin{equation}}
\def \ee  {\end{equation}}
\def \bex  {\begin{equation*}}
\def \eex  {\end{equation*}}
\def \bea {\begin{eqnarray}}
\def \eea {\end{eqnarray}}
\def \bal {\begin{align}}
\def \eal {\end{align}}

\newcommand{\R}{\ensuremath{\mathbb{R}}}

\allowdisplaybreaks
\title{\boldmath On quarkonium masses in 3D non-commutative space}

\author[a]{Benedek Bukor}
\author[b]{Juraj Tekel}

\affiliation[a]{Faculty of Mathematics, Physics and Informatics,\\ Comenius University, Mlynsk\'a Dolina, Bratislava, 842 48, Slovakia}
\affiliation[b]{Department of Theoretical Physics,\\ Faculty of Mathematics, Physics and Informatics,\\ Comenius University, Mlynsk\'a Dolina, Bratislava, 842 48, Slovakia
}

\emailAdd{bukor2@uniba.sk}
\emailAdd{juraj.tekel@fmph.uniba.sk}

\abstract{
We modify the calculation of quarkonium masses using the radial WKB and Pekeris-type approximations for the case of three-dimensional, rotationally invariant non-commutative space. We obtain corrections to the charmonium ($\text{c}\bar{\text{c}}$), bottomonium ($\text{b}\bar{\text{b}}$) and bottom-charmed meson ($\text{c}\bar{\text{b}}$) masses due to the discrete short distance structure of the space introduced by the space non-commutativity. For the fundamental length at the Planck scale we obtain relative correction at the order of $10^{-39}$, and taking into account the current experimental data, we obtain the upper bound at the order of ${10^{-18}\textrm{ m}}$ for the fundamental length scale of space.
}

\begin{document} 
\maketitle
\flushbottom


\section{Introduction}\label{sec_intro}

The contradiction between the observed stability of atoms and the instability of electromagnetically bound systems due the radiation of accelerating charges in classical electromagnetism has been solved by the uncertainty principle of quantum mechanics (QM) in 1920's \cite{griffits}. Since a particle cannot be localized arbitrarily precisely in the phase space, this space breaks down to a set of Planck cells of finite volume -- the ground states of system must occupy at least this Planck cell and not a smaller region. A similar mechanism has been proposed to solve the apparent contradiction of the observed stability of the vacuum of quantum field theory and its instability due to large space-time fluctuations in a naive version of quantum gravity \cite{doplicher}.

In both cases, we arrive at a notion of a non-commutative (NC) space \cite{cones}. Opposing to the usual manifold, coordinates in such spaces are described by non-commuting operators. This leads to an algebra of functions on NC space, which is non-commutative. The simplest example is the two-dimensional NC plane, a.k.a. the above mentioned phase space of QM particle in one dimension. Even simpler example are fuzzy spaces, special cases of NC spaces where the algebra of functions is finite dimensional, and thus can be viewed as matrix algebra --- the most prominent example being the fuzzy sphere \cite{sF20,sF21}. Formulation of quantum mechanics on NC and fuzzy spaces is quite straightforward, we can define the NC versions of all the relevant operators and then look for the solutions of the corresponding Schrödinger equation, for some of the early results see \cite{NCQM1,NCQM2}.

The general construction of non-commutative spaces usually follows the procedure of quantization of Poisson manifolds \cite{steinacker_review}. This works, however, only for even dimensional manifolds, and much more work is required to describe a space with three dimensions. Odd dimensional fuzzy spheres have been constructed in \cite{odd}, we will follow the construction of a three dimensional space of \cite{kovacik}. The next section gives technical details about the construction. Here, let us give a more intuitive picture. The space is constructed as an infinite set of fuzzy spheres with growing radius, which form a layered structure one could call a fuzzy onion. This leads to a different discreteness in the radial coordinate than in the angular coordinates, since the latter is smooth and still possesses the rotational symmetry while the former is rigid. However, the distance of the fuzzy-sphere-layers is the same as the characteristic distance of the NC structure of each layer, and thus the discreteness of the space is compatible throughout the whole 3D space.

In this paper, we will concentrate on quantum mechanical systems in such a space, with a plan of extracting at least some consequences of the non-trivial structure on their properties. Previously, hydrogen atom \cite{NCH}, particle in a potential well \cite{NCwell} and other properties of this space and related physics \cite{other1,other2,other3} have been considered, a more field theory oriented construction has been presented in \cite{Hammou:2001cc,other4}, and the idea has also appeared as a solution for a non-commutative black hole in \cite{Schupp:2009pt}. The system of our choice will be the bound states of two heavy quarks -- quarkonium \cite{quarkonium}. Such system is heuristically described by Cornell potential, which includes a Coulomb-like interaction between the quarks and attractive linear part, which describes the strong interaction leading to the QCD confinement. For low energy bound states of the heavy quarks the classical speeds are rather small, and one can use non-relativistic quantum mechanics to describe such system \cite{AEIM,NUM,AIM,LTM,ANN}. Such states have been observed in particle accelerators, and masses of several of them are known \cite{pdg2}. Some attempts to treat quarkonium in NC space have been done in \cite{NC_bb,NC_cc}.

Quarkonium states are especially promising candidates to test the quantum structure of space-time, since they represent a system with reasonably small characteristic distance. The two quarks orbit quite close to each other and thus probe any new features of the space-time much better than particles in different, larger systems -- e.g. the electrons in atoms.

This paper is organized as follows. In the preliminary section \ref{sec2}, we review the basic notions we will employ in the rest of the paper -- WKB approximation in non-relativistic QM, construction and properties of 3D NC space, calculation of quarkonium masses using the Cornell potential. Then we solve the analogous problem in the 3D NC space in section \ref{sec3}, and in section \ref{sec4}, we derive the NC corrections to the masses of quarkonium states.


\section{Preliminaries}\label{sec2}

\subsection{Quarkonium and quarkonium masses}

In this work, we will deal with the masses of bound states of two heavy quarks. The analytical treatment is rather phenomenological and the standard approach goes as follows.

Describe the two quarks of masses $m_1$ and $m_2$ as point masses with the Cornell potential
\begin{align}
    V_Q(r)=-\frac{C}{r}+B\,r\ .\label{cornpot}
\end{align}
The linear part is responsible for the quark confinement and the $r^{-1}$ part describes the electrostatic-like interaction between charged quarks. Both $B$ and $C$ could be in principle determined from the first principles and the underlying fundamental theory, but we will treat them as free parameters to be fixed by experimental data. More complicated potentials have been proposed (e.g. Killingback potential ${V_K(r)=-C/r+Br+Gr^2}$), but we will keep this simple form to have as few free parameters as possible. We will, however, keep the parameters $B$ and $C$ different for different composition of quarkonium states.

For large masses $m_1$ and $m_2$, the system will be well described as a non-relativistic two body problem in QM\footnote{This is an assumption to be checked {\emph{a posteriori}}, we will get back to this is section \ref{sec4}.}.The binding energies -- which will be a function of two quantum numbers $n$ and $l$ -- are thus given by the Schrödinger equation for a particle of mass ${\mu=m_1 m_2/(m_1+m_2)}$ in the potential $V_Q(r)$. The observed mass of the bound state is then\footnote{We will be mostly working with physical quantities in natural units. The usage of the SI units will be emphasized or should be clear from the context.} 
\begin{align}
   {\color{black}  M_{nl}=m_1+m_2+E_{nl}\ .} \label{Mnl}
\end{align}
A significant part of our work will thus be devoted to the determination of the binding energies $E_{nl}$. On one hand, we will look for corrections due to the NC structure of the space, but we will also present a modified way to treat the QM problem with the Cornell potential in the classical space, leading to some new results even with no non-commutativity.

\subsection{Non-relativistic quantum mechanics}\label{sec2sub1}

\color{black}
\subsubsection*{Standard Schrödinger equation}
Our starting point, the time-independent Schrödinger equation, is
\begin{equation}
\left[-\frac{\hbar^2}{2\mu}\Delta+V(r)\right]\Psi(\textbf{r})=E\Psi(\textbf{r})\ ,
\end{equation}
with the 3D position vector $\textbf{r}$ and ${r=|\textbf{r}|}$. All of the systems which we are going to deal with -- the Cornell potential $V_Q(r)$, the Killingback potential $V_K(r)$ and the Coulomb potential ${V_C(r)=-e^2/\lr{4\pi\varepsilon_0 r}}$ -- are rotationally invariant, and thus are also the respective Hamiltonian operators. Hence, the solution is reasonably sought in a separated form
\begin{equation}
\Psi_{lm}(\textbf{r})=R_{l}(r)Y_{lm}(\vartheta, \varphi)\ ,\label{varsep}
\end{equation}
where $l$ is the quantum number corresponding to the angular momentum, $m$ is the quantum number corresponding to the third component of angular momentum and $Y_{lm}(\vartheta, \varphi)$ is the standard spherical harmonic function. 
The $r$-dependent function $R_l(r)$ satisfies the radial Schrödinger equation\footnote{In particular cases, it will be more convenient to tackle the radial wave function $R_{l}(r)$ in a form $R_{l}(r)=K_{l}(r)r^l$ where the term $r^l$ describes the behaviour of $R_{l}(r\rightarrow 0)$.} 
\begin{align}{\color{black}
0=R''(r)+\frac{2}{r}R'(r)+
\frac{2\mu}{\hbar^2}\left(E-V(r)-\frac{\hbar^2}{2\mu}\frac{l(l+1)}{r^2}\right)R(r)\ .\label{radialschr}}
\end{align}

In the case of the Coulomb potential $V_C(r)$ with SI units, the equation of the radial part is
\begin{equation}
R''(\rho)+\frac{2}{\rho}R'(\rho)+\left(-\epsilon+\frac{2}{\rho}-\frac{l(l+1)}{\rho^2}\right)R(\rho)=0\label{rad1}\ ,
\end{equation}
{\color{black} where ${\epsilon=E/\lr{-\frac12\mu c^2\alpha^2}}$ is the dimensionless energy, ${\rho=r/a_B}$ is the dimensionless radial coordinate, ${a_B=\hbar/\lr{\mu c\alpha}}$ is the Bohr radius, ${\alpha=e^2/\lr{4\pi\varepsilon_0\hbar c}}$ is the fine-structure constant and in the present case ${\mu=m_e m_p/(m_e+m_p)\approx m_e}$ is the mass of the electron.}
The formula for the discrete energy levels is
\begin{equation}{\color{black}
E_N=-\frac12m_e c^2\alpha^2\frac{1}{N^2}\ ,\ N=l+1,\ l+2,\ \ldots\label{EN}\ ;}
\end{equation}
for the complete solution of \eqref{rad1}, see \cite{griffits}.

In the case of the Cornell potential $V_Q(r)$, the equation of the radial part is
\begin{equation}
R''(\zeta)+\frac{2}{\zeta}R'(\zeta)+\left(\epsilon+\frac{c}{\zeta}-b\zeta-\frac{l(l+1)}{\zeta^2}\right)R(\zeta)=0\label{rad2}\ ,
\end{equation}
 where ${\zeta=r/r_Q}$ is the dimensionless radial coordinate, $r_Q=\sqrt{C/B}$ is the typical distance of the Cornell potential\footnote{Some studies, e.g. \cite{nigeria} suggest to consider the typical distance as an independent parameter from the potential itself, but we will express it with the help of the other parameters of the potential. According to the dimensional analysis there exist multiple different typical distances, our choice is based on the fact that only for this distance $r_Q$ is the equation ${V_Q(r_Q)=0}$ valid.}, ${\epsilon=2\mu Er_Q^2/\hbar^2}$ is the dimensionless energy,  ${c=2\mu Cr_Q/\hbar^2}$ is the dimensionless Coulombic part\footnote{In those few cases where the SI units are adopted, the speed of the light is denoted $c$. In those cases where the dimensionless Coulombic part $c$ is used, we will not adopt the SI units. Thus, the two different interpretations of the denotation $c$ will be perceptible.} and ${b=2\mu Br_Q^3/\hbar^2}$ is the dimensionless linear confinement term. For the charmonium and the bottomonium systems ${\mu=m_q m_{\bar{q}}/(m_q+m_{\bar{q}})=m_q/2}$ where ${m_q=m_{\bar{q}}}$ is the mass of the considered (anti)quark. Equation \eqref{rad2} cannot be solved exactly, there exist only approximation methods for the energy spectrum of ${V_Q(r)}$: analytical exact iteration method \cite{AEIM}, Nikiforov-Uvarov method \cite{NUM},  asymptotic iteration method \cite{AIM}, Laplace transformation method \cite{LTM}, method using artificial neural networks \cite{ANN}.
 
We will use a different method, based on WKB and Pekeris-type approximations\footnote{We use terminology where Pekeris approximation refers to the expansion of the centrifugal term of the effective potential, while Pekeris-type approximation refers to expansion of other terms in $V(r)$ \cite{terminology}.}, based on \cite{svk} and \cite{nigeria}.

\subsubsection*{Radial WKB approximation}
The WKB method is a technique for obtaining approximate solution to the time-independent Schrödinger equation, see \cite{WKB1, griffits}.

In the ‘1D' WKB approximation, the wave function $\psi(x)$ is defined for ${x\in\mathbb{R}}$, whereas for rotationally invariant potentials the radial wave function $R(r)$ is defined for  ${r\in\mathbb{R}^+}$; 
hence the ‘standard' WKB method directly cannot be used.
To avoid this problem, region ${0< r<\infty}$ can be mapped to ${-\infty< x<\infty}$ by the bijective transformation\footnote{Since $x$ is an auxiliary variable, we will not pay attention to its physical dimension.} ${r=e^x}$. The radial wave function is being searched for in the form 
\begin{align}
R\left(r=e^x\right)\rightarrow R(x)=U(x)e^{f(x)}\label{Rfx}\ ,
\end{align}
where the function $f(x)$ is chosen such that the first derivative of $U(x)$ disappears from the radial Schrödinger equation \eqref{radialschr}. In the present case, ${f(x)=-x/2}$, thus we get
\begin{align}{\color{black} 
0=U''(x)+\frac{e^{2x}}{\hbar^2}\underbrace{2\mu\left(E-V(e^{x})-\frac{(l+\frac12)^2\hbar^2}{2\mu e^{2x}}\right)}_{p^2(e^x)}U(x)\ .}
\end{align}
Hence, the particle's semi-classical momentum $p(r)$ can be picked out:
\begin{equation}{\color{black} 
p(r)=\sqrt{2\mu\left(E-V(r)-\frac{(l+\frac12)^2\hbar^2}{2\mu r^{2}}\right)}\ ,\label{pr}}
\end{equation}
to be substituted to the WKB condition
\begin{equation}
\frac{1}{\hbar}\int^{r_2}_{r_1}\mathrm{d}{r}\,p(r)=\left(n+\frac12\right)\pi\ , \ n\in \mathbb{Z}^{+}_{0}\ ,\label{WKB}
\end{equation}
where the limits of the integration ${\{r_1, r_2;\ 0<r_1<r_2\}}$ 
can be found utilizing the relation ${p(r_1)=p(r_2)=0}$. Equation \eqref{WKB} is a condition on energy levels $E$.

\color{black}
\subsubsection*{Energy spectrum of the Cornell potential}
We would like to apply the WKB condition \eqref{WKB} for the potential described by the mathematical formula ${V_K(r)=-C/r+Br+Gr^2}$. This means to find the energies determined by
\begin{equation}{\color{black} 
\frac{1}{\hbar}\int^{r_2}_{r_1}\mathrm{d}{r}\sqrt{2\mu\left(E-\lr{-\frac{C}{r}+Br+Gr^2}-\frac{(l+\frac12)^2\hbar^2}{2\mu r^{2}}\right)}=\left(n+\frac12\right)\pi\ , \ n\in \mathbb{Z}^{+}_{0}\ ,\label{WKBcorn}}
\end{equation}
for a given $n$ and $l$.
Here, we mostly follow \cite{nigeria}.
The investigation of such a more general potential than \eqref{cornpot} will be fruitful since in our further calculations we will get a similar WKB condition to this, see \eqref{pquark}. For our purposes it is just a secondary remark that $V_K(r)$ has the form of the Killingback potential, which is created by the merger of the Cornell potential $V_Q(r)$ and the potential of the isotropic linear harmonic oscillator $Gr^2$.

Using the dimensionless physical quantities, we get the previous equation in the form
\begin{equation}
\int^{\zeta_2}_{\zeta_1}\mathrm{d}{\zeta}\sqrt{\epsilon+\frac{c}{\zeta}-b\zeta-g\zeta^{2}-\frac{(l+\frac12)^2}{\zeta^{2}}}=\left(n+\frac12\right)\pi\ , \ 0<\zeta_1<\zeta<\zeta_2\ ,\label{WKBcorndim}
\end{equation}
where ${g=2\mu Gr_Q^4/\hbar^2}$ is the dimensionless quadratic term and the limits of the integration ${\{\zeta_1, \zeta_2\}}$ can be found as the positive roots of the expression under the square root on the LHS in \eqref{WKBcorndim}.
We substitute ${y=1/\zeta}$, so the previous equation becomes
\begin{equation}{\color{black} 
\int^{\zeta_1^{-1}}_{\zeta_2^{-1}}\frac{\mathrm{d}{y}}{y^2}\sqrt{\epsilon+cy-\frac{b}{y}-\frac{g}{y^2}-\lr{l+\frac12}^2y^2}=\left(n+\frac12\right)\pi\ . \label{WKBcorny}}
\end{equation}
To solve this integral, we use the Pekeris-type approximation, see \cite{pekeris,pekeris2,terminology}. The original idea is based on the expansion of the terms $b/y$ and $g/y^2$ in power series around $1$ in $y$-space which corresponds to the expansion around the dimensionful quantity ${1/r_Q}$ in $(1/r)$-space, see \cite{baku}. Originally \cite{nigeria}, the expansion point was left as a free parameter, but here, we chose it to be the characteristic distance of the Cornell potential $r_Q$. This is the subtle but important point where our approach differs from previous works.

The expansion is now done for the new variable ${z=y-1}$ around ${z=0}$  up to the $2^{\text{nd}}$ order of $z$ in the following way:
\begin{align}
\frac{b}{y}=b\lr{1+z}^{-1}=b\slr{1-z+z^2+\mathcal{O}\lr{z^3}}\approx b\lr{3-3y+y^2} ,\label{}
\end{align}
analogously,
\begin{align}
\frac{g}{y^2}=g\lr{1+z}^{-2}=g\slr{1-2z+3z^2+\mathcal{O}\lr{z^3}}\approx g\lr{6-8y+3y^2}\ .\label{}
\end{align}
The presented expansions are valid only for ${z=\varepsilon\ll1}$ which means we must consider dimensionless radii\footnote{This is an assumption to be checked {\emph{a posteriori}}, we will get back to this is section \ref{sec4}.} ${\zeta=1/(1+\varepsilon)\approx1-\varepsilon\approx1}$ in \eqref{WKBcorndim}, thus ${r\approx r_Q}$ in \eqref{WKBcorn}. 
On the one hand for larger quantum numbers $l$ the centrifugal term overshadows the manipulations with the terms $b/y$ and $g/y^2$, so the Pekeris-type approximation may work better for larger quantum numbers $l$. However, for larger quantum numbers $l$ the quarks may orbit in a greater radius than the typical distance $r_Q$, so the considered interval of $y$ may be ${0< y<1}$, thus the terms $b/y$ and $g/y^2$ may become dominant resulting in the misfunction of the Pekeris-type approximation. As we will see in section \ref{sec4}, it turns out that despite the present alarm, for larger quantum numbers $l$ our theoretical results are really in a better correspondence with the experimental values.

With the help of the this expansions, the equation \eqref{WKBcorny} becomes
\begin{align}
\sqrt{b+3g+\lr{l+\frac12}^2}&\int^{y_2}_{y_1}\frac{\mathrm{d}{y}}{y^2}\sqrt{-\frac{-\epsilon+3b+6g}{b+3g+\lr{l+\frac12}^2}+\frac{c+3b+8g}{b+3g+\lr{l+\frac12}^2}y-y^2}=\nonumber \\ 
&=\left(n+\frac12\right)\pi\ , \ \ 0<y_1<y<y_2\ , \label{viete}
\end{align}
and the limits of the integration ${\{y_1, y_2\}}$ can be found as the positive roots of the quadratic expression under the square root on the LHS in \eqref{viete}. For further calculations we take the advantage of the following integral:
\begin{equation}
\int^{y_2}_{y_1}\frac{\mathrm{d}{y}}{y^2}\sqrt{\lr{y_2-y}\lr{y-y_1}}=\pi\frac{\frac12\lr{y_1+y_2}-\sqrt{y_1y_2}}{\sqrt{y_1y_2}}\ ,\ 0<y_1<y<y_2\ .
\end{equation}
By the application of the Vieta's formulas for the quadratic expression under the square root in \eqref{viete},
we get
\begin{equation}
\sqrt{b+3g+\lr{l+\frac12}^2}\pi\frac{\frac12\lr{\frac{c+3b+8g}{b+3g+\lr{l+\frac12}^2}}-\sqrt{\frac{-\epsilon+3b+6g}{b+3g+\lr{l+\frac12}^2}}}{\sqrt{\frac{-\epsilon+3b+6g}{b+3g+\lr{l+\frac12}^2}}}=\left(n+\frac12\right)\pi\ .\label{dokopy}
\end{equation}
Using a little algebra, the dimensionless energy can be expressed from the previous equation:
\begin{equation} 
\lr{\epsilon}_{\text{WKB, PEK}}=-\frac14\slr{\frac{c+3b+8g}{n+\frac12+\sqrt{b+3g+\lr{l+\frac12}^2}}}^2+3b+6g\ .\label{ekillingback}
\end{equation}
Switching back to quantities with dimension, we get
\begin{equation}
\lr{E}_{\text{WKB, PEK}}=-\frac{\mu}{2\hbar^2}\slr{\frac{C+3Br_Q^2+8Gr_Q^3}{n+\frac12+\sqrt{\frac{2\mu}{\hbar^2}\lr{Br_Q^3+3Gr_Q^4}+\lr{l+\frac12}^2}}}^2+3Br_Q+6Gr_Q^2\ .\label{Ekillingback}
\end{equation}
Although it was not our primary target, we recovered the energy spectrum of the Killingback potential using radial WKB method and Pekeris-type approximation. Taking ${G=0}$ in \eqref{Ekillingback} and substituting our formula for $r_Q$, we obtain the approximate energy spectrum of the Cornell potential:
\begin{equation}
\lr{E_{nl}}_{\text{WKB, PEK}}=-\frac{2\mu}{\hbar^2}\slr{\frac{2C}{n+\frac12+\sqrt{\frac{2\mu}{\hbar^2}C\sqrt{\frac{C}{B}}+\lr{l+\frac12}^2}}}^2+3\sqrt{BC}\ ;\label{Ecornell}
\end{equation}
whose dimensionless version is obtained by taking ${g=0}$ in \eqref{ekillingback}:
\begin{equation} 
\lr{\epsilon_{nl}}_{\text{WKB, PEK}}=-\frac14\slr{\frac{c+3b}{n+\frac12+\sqrt{b+\lr{l+\frac12}^2}}}^2+3b\ .\label{ecornell}
\end{equation}
\color{black} 
Taking ${G=B=0}$, ${r_Q=a_B}$ and ${C=e^2/\lr{4\pi\varepsilon_0}}$ in \eqref{Ekillingback} and respectively introducing a new quantum number ${N\equiv (n+l+1) \in  \mathbb{N}}$, we shall write the formula of the approximate energy spectrum of the Coulomb potential with SI units:
\begin{equation}
\lr{E_{N}}_{\text{WKB}}=-\frac{m_e}{2\hbar^2}\lr{\frac{e^2}{4\pi\varepsilon_0N}}^2=-\frac{1}{2}m_e\alpha^2c^2\frac{1}{N^2}=E_{N}\ .\label{}
\end{equation}
Since we used the Pekeris-type approximation for terms directly proportional to $B$ and $G$, taking them zero, the Pekeris-type approximation does not effect the energy spectrum of the bound states of electron of mass $m_e$ in the hydrogen atom.
However, we must admit that we recovered the exact energy spectrum by the merest chance.
It cannot be forgotten that the WKB method works the best for large momentum, small de Broglie wavelength $\left(\mathrm{d}{\lambda_{\text{de Broglie}}}/\mathrm{d}{x}\ll 2\pi\right)$ and hence for large quantum numbers: $n$, $l$ and $N$, see \cite{tong}. 

\color{black}

\subsection{Three dimensional non-commutative space and NC QM}

\subsubsection*{Construction of the space}

In this section, we will review the construction of a 3D noncommutative space based on \cite{kovacik}. In standard QM physical quantities are represented by hermitian operators acting on the states of the Hilbert space $\mathcal{H}$ -- on the wave functions. In QM with NC coordinates this idea is still preserved, but the NC wave functions are composed of operators acting in the auxiliary Fock space. The NC space $\R^{3}_{\lambda}$ can be described by a model of concentric fuzzy spheres with increasing radius. The commutator of coordinates is defined as
\begin{equation}
\left[x_i, x_j\right]=2i\lambda\varepsilon_{ijk}x_k\ , \label{comx}
\end{equation}
where the parameter $\lambda$ describes the fuzziness of the space structure\footnote{Note that we would have a similar looking commutation relation for a single fuzzy sphere. The difference is that in that case, operator \eqref{oper} would be restricted to have a fixed value, restricting us onto a subset of the Fock space. Here, we allow any value of $r$, leading to spheres of all radii.}. The Fock space is accompanied by two sets of creation $a^{\dag}$ and the annihilation $a$ operators; their commutation relations are
\begin{equation}
\left[a_\alpha, a_\beta^{\dag}\right]=\delta_{\alpha\beta}\ ,\ \Big[a_\alpha, a_\beta\Big]=\left[a_\alpha^{\dag}, a_\beta^{\dag}\right]=0\ ,\ \alpha,\beta\in\{1,2\}\ .
\end{equation}

The appealing fact is that all the operators of all the relevant physical quantities can be constructed with this choice of the operators $a^{\dag}$ and $a$. The definition of position operator\footnote{The operators acting on the states of the Fock space are marked without hat.}
\begin{equation}
x_j=\lambda\sigma^{j}_{\alpha\beta}a_\alpha^{\dag}a_\beta\ ,\ j\in\{1,2,3\} \
\end{equation}
obeys commutator \eqref{comx}, $\sigma^{j}$ is the corresponding Pauli matrix. The magnitude of the position vector is defined as 
\begin{equation}
r=\lambda\left(a^{\dag}_\alpha a_\alpha+1\right)\ ;\label{oper}
\end{equation}
for further calculations we take the advantage of the operator
\begin{equation}
\tilde r=\lambda\left(a_\alpha^{\dag}a_\alpha\right)\ .\label{tilder}
\end{equation}

The wave functions $\Psi(x_j)$ can be expressed instead of $x_j$ in terms of $a^{\dag}$ and $a$ in the way:
\begin{equation}
\Psi=\sum C_{m_1m_2n_1n_2}(a_1^{\dag})^{m_1}(a_2^{\dag})^{m_2}(a_1)^{n_1}(a_2)^{n_2}\ . \label{Psi}
\end{equation}
The number of $a^{\dag}$'s and $a$'s are equal in the definition of $x_j$, thus they must be equal in the expression for $\Psi$, too, so ${m_1+m_2=n_1+n_2}$, where ${m_1,m_2,n_1,n_2\in\mathbb{Z}^{+}_{0}}$. Moreover, for the arrangement of operators $a^{\dag}$ and $a$, the normal ordering is imposed by hand, i.e. all the creation operators $a^\dagger$ are to the left of any annihilation operators $a$.

The NC analogy of the method of separation the variables in standard QM \eqref{varsep} leads to the separated form
\begin{equation}
\Psi_{lm}=\lambda^l\sum_{(lm)}\frac{(a^{\dag}_1)^{m_1}(a^{\dag}_2)^{m_2}}{m_1!m_2!}\colon \mathcal{K}_l(\tilde r)\colon\frac{a_1^{n_1}\left(-a_2\right)^{n_2}}{n_1!n_2!}\ , \label{eigen}
\end{equation}
where $\mathcal{K}_l(\tilde r)$ is the NC analogy of the radial wave function $K_l(r)$, constructed in the same way as \eqref{Psi}, and between the colon marks normal ordering is needed. The summation goes over all quantum numbers ${l=m_1+m_2=n_1+n_2}$ and ${m=1/2\ (m_1-m_2-n_1+n_2)}$, which will play the same role as quantum numbers in \eqref{varsep}.

The action of the operator $\hat r$ from \eqref{oper} on the wave function\footnote{The operators acting on the states of the Hilbert space $\mathcal{H}_{\lambda}$ are marked with hat.} in \eqref{eigen} can be  calculated:
\begin{equation}
\hat r \Psi_{lm}=\lambda^l\sum_{(lm)}\frac{(a^{\dag}_1)^{m_1}(a^{\dag}_2)^{m_2}}{m_1!m_2!}
 \colon \left[(\tilde r+\lambda l+\lambda)\mathcal{K}+\lambda\tilde r \mathcal{K}'\right]\colon\frac{a_1^{n_1}\left(-a_2\right)^{n_2}}{n_1!n_2!}\ .\label{r}
\end{equation}

The definition of the Laplace operator in $\mathcal{H}_{\lambda}$ is\footnote{Expression $\frac{1}{r}$ is understood as the inverse operator to the operator $\hat r$.} 
\begin{equation}
\hat \Delta_{\lambda} \Psi=-\frac{1}{\lambda r}\left[{\hat a_{\alpha}^{\dag}}, \left[{\hat a_{\alpha}, \Psi}\right]\right]\ ;
\end{equation}
the action of the operator ${\big[{\hat a_{\alpha}^{\dag}}, \left[{\hat a_{\alpha},\ \cdot\ }\right]\big]}$ on the wave function\footnote{Note that in this case the creation and annihilation operators are rewarded with hat since in the present case they act on the wave function. We want to stress that marking the operator with hat is independent from the operator itself, the decisive factor is whether the operator is understood as acting on the states of the Fock space or the Hilbert space $\mathcal{H}_{\lambda}$ .} in \eqref{eigen} is
\begin{align}
&\left[{\hat a_{\alpha}^{\dag}}, \left[{\hat a_{\alpha}, \Psi_{lm}}\right]\right]=\lambda^l\sum_{(lm)}\frac{(a^{\dag}_1)^{m_1}(a^{\dag}_2)^{m_2}}{m_1!m_2!} \colon \left[-\lambda\tilde r \mathcal{K}''-2(l+1)\lambda \mathcal{K}'\right]\colon\frac{a_1^{n_1}\left(-a_2\right)^{n_2}}{n_1!n_2!}\ ,\label{aa}
\end{align}
where $\mathcal{K}\equiv \mathcal{K}_{l}(\tilde r)$ and similarly for derivatives.

A different construction of a 3D non-commutative space has been presented in \cite{kupriyanov}. While it preserves the symmetries of the space, the Hamiltonian is constructed in a way that breaks the azimuthal symmetry of the problem, leading to results different from the ones presented here. In this framework, the hydrogen atom \cite{kupriyanov_H} has been considered, and recently also the spectrum of the $\text{b}\bar{\text{b}}$ \cite{NC_bb} and $\text{c}\bar{\text{c}}$ \cite{NC_cc} states. We will comment on these results where appropriate.

\subsubsection*{NC Schrödinger equation}

Putting together our knowledge of the NC potential and the NC Laplace operator, we acquire the Hamiltonian operator $\hat H_{\lambda}$ in the form
\begin{equation}{\color{black}
\hat H_{\lambda} \Psi=\left[-\frac{\hbar^2}{2\mu}\hat\Delta_{\lambda}+V(\hat r)\right]\Psi\ .\label{NCschr}}
\end{equation}
This leads to the eigenvalue problem for the Hamiltionan of the form
\begin{equation}{\color{black}
\frac{\hbar^2}{2\mu\lambda}\left[{\hat a_{\alpha}^{\dag}}, \left[{\hat a_{\alpha}, \Psi}\right]\right]+\hat r V(\hat r)\Psi=E\hat r \Psi\ .}
\end{equation}
Note that due to the presence of $\lambda$ in operator $\hat r$ on the RHS of the above equation and due to \eqref{r} the energy and the non-commutative length scale mix, and the effect of non-commutativity cannot be easily written as a small perturbation Hamiltonian. This means that the standard perturbation theory of quantum mechanics is not going to be applicable, and we will have to use other methods.

\subsubsection*{NC Hydrogen atom}

\color{black}Based on \eqref{NCschr} the NC analog of the Schrödinger equation with the Coulomb potential $V_C(\hat r)$ in $\mathbb{R}^{3}_{\lambda}$ is 
\begin{equation}{\color{black}
\frac{\hbar^2}{2m_e\lambda r}\left[{\hat a_{\alpha}^{\dag}}, \left[{\hat a_{\alpha}, \Psi}\right]\right]-\frac{1}{4\pi\varepsilon_0}\frac{e^2}{r}\Psi=E\Psi\ .}
\end{equation}
Multiplying the previous equation with $\hat r$ and inserting \eqref{aa} and \eqref{r}, we get
\begin{align}
\colon\tilde \rho \mathcal{K}''&+[-\epsilon\sigma\tilde \rho+2(l+1)]\mathcal{K}'+[-\epsilon\tilde \rho-\epsilon\sigma(l+1)+2]\mathcal{K}\colon=0\ ,\label{schrK}
\end{align}
where $\tilde \rho$ is the dimensionless operator $\tilde r$ from \eqref{tilder} and ${\sigma=\lambda/a_B}$; this equation is the analog of the radial Schrödinger equation in NCQM expressed with $\mathcal{K}(\tilde \rho)$ instead of $\mathcal{R}(\tilde \rho)$. 
Equation \eqref{schrK} can be interpreted in the way that the configuration of the operators on the left side results in a zero operator. In the case we remove the colon marks, we get the differential equation
\begin{align}
\rho K''&+[-\epsilon\sigma\rho+2(l+1)]{K}'+[-\epsilon\rho-\epsilon\sigma(l+1)+2]{K}=0\ ,\label{schrK1}
\end{align}
for an ordinary function $K$ of an ordinary variable $\rho$. The claim is that one obtains solution to \eqref{schrK} by replacing $\rho$ in the solution to \eqref{schrK1} with operator $\tilde \rho$, see \cite{kovacik}.

The solution of \eqref{schrK1} for the radial wave function ${K(\rho)}$ results in hypergeometric function, and the exact energy spectrum of the bound states with SI units is
\begin{equation}
E^{\lambda}_N=\frac{\hbar^2}{m_e\lambda^2}\left(1-\sqrt{1+\frac{m_e^2c^2\alpha^2}{\hbar^2}\frac{\lambda^2}{N^2}}\right)\ .\label{presneen}
\end{equation}
One can see that the energy levels calculated in standard QM increase due to the effect of the NC space structure. Taking the limit ${\lambda \rightarrow 0}$ we recover the energy spectrum \eqref{EN}.
\color{black}


\section{Energy spectrum of Cornell potential in NC QM}\label{sec3}

\color{black}Based on \eqref{NCschr} the NC analog of the Schrödinger equation with the Cornell potential $V_Q(\hat r)$ in $\mathbb{R}^3_{\lambda}$ is 
\begin{equation}{\color{black}
\frac{\hbar^2}{2\mu\lambda r}\left[{\hat a_{\alpha}^{\dag}}, \left[{\hat a_{\alpha}, \Psi}\right]\right]+\left(-\frac{C}{r}+B\hat r\right)\Psi=E\Psi\ . \label{schrr3}}
\end{equation}
It is inevitable to calculate the action of operator $\hat r^2$ on the wave function $\Psi_{lm}$, with the help of \eqref{r}, we get
\begin{align}
\hat r^2\Psi_{lm}\equiv\ \hat r\left(\hat r\Psi_{lm}\right) \,=\,&\lambda^l\sum_{(lm)}\frac{(a^{\dag}_1)^{m_1}(a^{\dag}_2)^{m_2}}{m_1!m_2!}
\colon\Big[\tilde r^2\mathcal{K}+\lambda\Big((2l+3)\tilde r\mathcal{K}+2\tilde r^2\mathcal{K}'\Big)+ \nonumber \\&+
\lambda^2\Big(\left(l+1\right)^2 \mathcal{K}+\left(2l+3\right)\tilde r\mathcal{K}'+\tilde r^2\mathcal{K}''\Big)\Big]\colon\frac{a_1^{n_1}\left(-a_2\right)^{n_2}}{n_1!n_2!}\ .
\end{align}
By the introduction of the dimensionless parameter ${\sigma=\lambda/r_Q}$, equation {\eqref{schrr3}} yields the radial
Schrödinger equation in the NC configuration
space:
\color{black}
\begin{align}
R''+\frac{2}{\zeta}R'-\frac{l(l+1)}{\zeta^2}R&+\left(\frac{c}{\zeta}-b\zeta\right)R+\epsilon R+\sigma\left(\epsilon R'+\frac{\epsilon }{\zeta}R-2b\zeta R'-3b R\right)+\nonumber \\&
+\sigma^2\Big(-b\zeta R''-3b R'-\frac{b}{\zeta}R\Big)=0
\label{rad3}\ .
\end{align}
In the case we take the limit ${b\rightarrow0}$ in equation \eqref{rad3}, we recover the radial Schrödinger equation \eqref{schrK1} expressed for a special parameter $c$; in the case we take the limit ${\sigma\rightarrow0}$ in equation \eqref{rad3}, we recover the radial Schrödinger equation \eqref{rad2}.
Analogously to the described method in subsection {\ref{sec2sub1}}, we substitute ${\zeta=e^x}$ and find the proper $f(x)$ in \eqref{Rfx} such that $U'(x)$ term will not be present. We recover the previous equation in the form
\begin{align}
0=U''(x)+e^{2x}\underbrace{\frac{\epsilon+\lr{\frac{c}{e^x}-b e^x}-\frac{\left(l+\frac{1}{2}\right)^2}{e^{2 x}}-\sigma^2\left(c b-\frac{b}{e^x}\lr{l^2+l+\frac12}+\frac{\epsilon^2}{4}\right)}{\left(1-b \sigma^2 e^x\right)^2}}_{\kappa^2(e^x)}U(x)\ .\label{cornU}
\end{align}
The parameter $\sigma$ is deemed to be small, since $r_Q$ is expected to be at the order of the size of the quarkonium states\footnote{This is an assumption to be checked {\emph{a posteriori}}. We will return to the value of $r_Q$ in section \ref{sec4}.}, which is roughly $10^{-16}\,\text{m}$ \cite{size}, and $\lambda$ is expected to be at the Planck scale, i.e. $10^{-35}\,\text{m}$, thus ${\sigma\approx10^{-19}}$. For this reason the denominator in \eqref{cornU} is not expected to become zero. We do not cause a great deal of inaccuracy if all the terms in the $3^{\text{rd}}$, $4^{\text{th}}, \ldots$ orders of $\sigma$ become neglected, the most intrinsic information is still hidden in terms up to the $2^{\text{nd}}$ order of $\sigma$. Keeping this in mind, we acquire the dimensionless semi-classical momentum $\kappa(\zeta)$ as
\begin{align}
\kappa(\zeta)=
\sqrt{\epsilon+\left(\frac{c}{\zeta}-b \zeta\right)-\frac{\lr{l+\frac12}^2}{\zeta^{2}}+\sigma^2\left(c b+2\epsilon b \zeta-2 b^2 \zeta^{2 }-\frac{b}{\zeta}l(l+1)-\frac{\epsilon^2}{4}\right)}+\mathcal{O}\lr{\sigma^3}
\ .
\end{align}
Hence, the WKB condition is attained considering all the terms up to the $2^{\text{nd}}$ order of $\sigma$:
\begin{equation}
\left(n+\frac12\right)\pi\approx\int_{\zeta_1}^{\zeta_2}\mathrm{d}{\zeta}\sqrt{
\begin{array}{l}
\vspace{0.2cm}
\slr{\epsilon+\sigma^2\lr{c b-\frac{\epsilon^2}{4}}}+\slr{c-\sigma^2bl(l+1)}\frac{1}{\zeta}-\\
\ \ -
b\lr{1-2\sigma^2\epsilon}\zeta-2\sigma^2b^2\zeta^2-\frac{\lr{l+\frac12}^2}{\zeta^2}
\end{array}}\ . \label{pquark}
\end{equation}
\color{black} 
As we have already become aware, the form of the expression under the square root reminds us of the WKB condition \eqref{WKBcorndim}. Although, in this case, the coefficients standing before the different powers of $\zeta$ contain the effect of the blur of the space $\sigma$. 
As a solution to \eqref{pquark}, now we can utilize the derived energy spectrum of the Killingback potential. Substituting the proper coefficients to the relation \eqref{ekillingback}, we get the equation up to the $2^{\text{nd}}$ order of $\sigma$
\begin{align}
\epsilon&+\sigma^2\lr{c b-\frac{\epsilon^2 }{4}}\approx\lr{-\frac14\slr{\frac{c+3b}{n+\frac12+\sqrt{b+\lr{l+\frac12}^2}}}^2+3b}+\sigma^2 \left[12 b^2-6 b \epsilon+\vphantom{\frac{ (c+3 b)^2 \left(3 b^2-b \epsilon\right)}{\sqrt{b+\lr{l+\frac12}^2} \left(n+\frac12+\sqrt{b+\lr{l+\frac12}^2}\right)^3}}\right. \nonumber\\
&\left.+ \frac{ (c+3 b)^2 \left(3 b^2-b \epsilon\right)}{2\sqrt{b+\lr{l+\frac12}^2} \left(n+\frac12+\sqrt{b+\lr{l+\frac12}^2}\right)^3}-\frac{ b(c+3 b) \left(16 b-6 \epsilon-l (l+1)\right)}{2\left(n+\frac12+\sqrt{b+\lr{l+\frac12}^2}\right)^2}\vphantom{\frac{1}{2}}\right]\ ,
\end{align}
where now we can clearly see the mentioned mixture of the energy and the non-commutative length scale. This equation is a quadratic one for the variable $\epsilon$, the solution contains two possible roots. The first one does not satisfy the desired property that for ${\sigma\rightarrow0}$ it gives back the initial energy spectrum \eqref{ecornell}, so it is abandoned. Considering all the terms up to the $2^{\text{nd}}$ order of $\sigma$, the second root is 

\begin{align}
\lr{\epsilon^{\sigma}_{nl}}_{\text{WKB, PEK}}\, &= \,\lr{\epsilon_{nl}}_{\text{WKB, PEK}}+\sigma^2\lr{\epsilon^{(2)}_{nl}}_{\text{WKB, PEK}}+\dots=\nonumber\\ 
&=-\frac14\slr{\frac{c+3b}{n+\frac12+\sqrt{b+\lr{l+\frac12}^2}}}^2+3b+\sigma^2\left(\frac{b \left(105 b^2+62 b c+9 c^2\right)}{8\left[n+\frac12+\sqrt{b+( l+\frac12)^2}\right]^2}+\vphantom{\frac{b (c+3 b)^4}{8\sqrt{b+( l+\frac12)^2} \left[n+\frac12+\sqrt{b+( l+\frac12)^2}\right]^5}+\frac{ b(c+3 b)l (l+1)}{2\left(n+\frac12+\sqrt{b+( l+\frac12)^2}\right)^2}}\right. \nonumber\\ 
&+\frac{b (c+3 b)^4}{8\sqrt{b+( l+\frac12)^2} \left[n+\frac12+\sqrt{b+( l+\frac12)^2}\right]^5}+\frac{ b(c+3 b)l (l+1)}{2\left[n+\frac12+\sqrt{b+( l+\frac12)^2}\right]^2}-\nonumber\\ 
&\left.-\frac{(45 b-c) (c+3 b)^3}{64 \left[n+\frac12+\sqrt{b+( l+\frac12)^2}\right]^4}-\frac{b}{4}\lr{15b+4c}\vphantom{\frac{1}{2}}\right)+\dots \ .\label{emod}
\end{align}
Note that taking the limit ${b\rightarrow0}$, which means we talk about the hydrogen atom, all the terms vanish from the expression of the modification except one term:
\begin{equation}
   \lim_{b \to 0}\lr{\epsilon^{\sigma}_{nl}}_{\text{WKB, PEK}} = -\frac{c^2}{4N^2}+\sigma^2\frac{c^4}{64 N^4}+\dots\ ;
\end{equation}
this remained term is responsible for the modification of the energy spectrum of the hydrogen atom \eqref{presneen}. 
In the limit ${b\rightarrow0}$, the Pekeris-type approximation does not effect the non-commutative energy correction of the hydrogen atom, i.e. by the only approximation -- the radial WKB method -- we recovered the exactly same modification of the hydrogen atom's energy spectrum up to the $2^{\text{nd}}$ order of $\sigma$ as the exact solution \eqref{presneen} gives. In fact, the NC Coulomb potential can be solved exactly with the help of the WKB approximation, and the solution gives back the exact non-commutative energy spectrum \eqref{presneen}, see \cite{svk}.

As an interesting observation, note that the large $n$ and $l$ correction in \eqref{emod} is finite, opposing to the case of the isotropic harmonic oscillator \cite{svk}.

\color{black} 

\section{Charmonium, bottomonium and bottom-charmed meson masses}\label{sec4}

\subsection{Charmonium, bottomonium and bottom-charmed meson masses in classical space}
With the help of equations \eqref{Mnl} and \eqref{Ecornell}, we get the mass of the quarkonium system in the commutative space:
\begin{equation}{\color{black}
   M_{nl}=(m_1+m_2)-\frac{2\mu}{\hbar^2}\slr{\frac{2C}{n+\frac12+\sqrt{\frac{2\mu}{\hbar^2}C\sqrt{\frac{C}{B}}+\lr{l+\frac12}^2}}}^2+3\sqrt{BC}\ .\label{vsetko}}
\end{equation}
Until now we have treated $B$ and $C$ as free parameters, now we fix them with the help of the actual experimental data \cite{pdg1, pdg2}. We substitute the experimental data for 1S and 2S states to equation \eqref{vsetko}, hence we obtain the free parameters by solving the two equations numerically for the given system. From this moment on, we consider them as constants -- even though different for different quark composition of the states -- the mass of other excited states will be obtained and the results will be compared with the available experimental data in the case of the charmonium and bottomonium systems. 

\subsubsection*{Charmonium and bottomonium mass spectrum}

The results for the charmonium states are given by the following table.

\begin{center}
\begin{tabular}{||c |c|c |c||} 
 \hline\hline
 $\text{c}\bar{\text{c}}$ meson &  ${m_q=1.27\ \text{GeV}}$ & ${B=0.322\ \text{GeV}^2}$ & ${C=0.891}$\\ [1ex]
 \hline\hline
  state & particle & present work $M_{nl}$ [GeV] & experimental data $M_{nl}$ [GeV]\\ [1ex]
 \hline\hline
 1S & $J/\psi$(1S) & used for $B$, $C$ & 3.097  \\ 
 \hline
 2S & $\psi$(2S) & used for $B$, $C$ & 3.686 \\
 \hline
 3S & $\psi$(4040) & 3.889 & 4.039 \\
 \hline
 4S & $\psi$(4230) & 3.982 &  4.223 \\
 \hline
 1P & $\chi_{\text{C1}}$(1P) & 3.518 & 3.511  \\ 
 \hline
 2P & $\chi_{\text{C2}}$(3930) & 3.823 & 3.923  \\
 \hline
 1D & $\psi$(3770) & 3.787 & 3.774  \\
 \hline\hline
\end{tabular}
\end{center}


From the known states for the given values of $n$ and $l$ that differ by other quantum numbers, we use the same states as the authors of \cite{nigeria}. This is mainly motivated by the experimental accessibility of these states in particle colliders.

The results for the bottomonium states are given by the following table.


\begin{center}
\begin{tabular}{||c |c|c |c||} 
 \hline\hline
 $\text{b}\bar{\text{b}}$ meson &  ${m_q=4.18\ \text{GeV}}$ & ${B=1.266\ \text{GeV}^2}$ & ${C=0.344}$\\ [1ex]
 \hline\hline
  state & particle & present work $M_{nl}$ [GeV] & experimental data $M_{nl}$ [GeV]\\ [1ex]
 \hline\hline
 1S & $\varUpsilon$(1S) & used for $B$, $C$ & 9.460  \\ 
 \hline
 2S & $\varUpsilon$(2S) & used for $B$, $C$ & 10.023 \\
 \hline
 3S & $\varUpsilon$(3S) & 10.178 & 10.355 \\
 \hline
 4S & $\varUpsilon$(4S) & 10.242 & 10.579 \\
 \hline
 1P & $h_{b}$(1P) & 9.942 & 9.899  \\ 
 \hline
 2P & $h_{b}$(2P) & 10.150 & 10.260  \\
 \hline
 1D & $\varUpsilon_2$(1D) & 10.140 & 10.164 \\
 \hline\hline
\end{tabular}
\end{center}



Before we proceed further let us point out that our results for the masses are in good agreement with the experiment and stand their ground among other, arguably more sophisticated models \cite{AEIM,NUM,AIM,LTM,ANN}. It is interesting that such a simple model and approach lead to these rather precise numbers.

From equation \eqref{Mnl} one can see that relativistic approach is not required provided that ${(m_1+m_2)\gg E_{nl}}$.
The lightest quarkonium from our three systems, i.e. the $\text{c}\bar{\text{c}}$ mesons of mass ${2m_q=2.54\,\text{GeV}}$, even in the heaviest state 4S has binding energy ${E_{nl}=1.68\,\text{GeV}}$, so our requirement for the non-relativistic consideration is reasonably valid even in this most extreme case. For large masses $m_1$ and $m_2$, the system is indeed well described as a non-relativistic two body problem in QM, the relativistic corrections are negligible.



\subsubsection*{Bottom-charmed meson's mass spectrum}

Since the results of our model are in a satisfying correspondence with the experimental data, we applied our method to find the lowest masses of the bottom-charmed meson's states. Experimental data exist only for the states 1S and 2S, so the constants $C$ and $B$ can be still determined, but the theoretical masses of the higher states cannot be compared to the real masses and are a prediction of our model.

\begin{center}
\begin{tabular}{||c |c|c |c||} 
 \hline\hline
 $\text{c}\bar{\text{b}}$ meson &  ${\mu=0.97\ \text{GeV}}$ & ${B=0.604\ \text{GeV}^2}$ & ${C=0.603}$\\ [1ex]
 \hline\hline
  state & particle & present work $M_{nl}$ [GeV] & experimental data $M_{nl}$ [GeV]\\ [1ex]
 \hline\hline
 1S & $B_{c}^+$ & used for $B$, $C$ & 6.274  \\ 
 \hline
 2S & $B_{c}^{\pm}$(2S) & used for $B$, $C$ & 6.871 \\
 \hline
 3S & -- & 7.054 & no data \\
 \hline
 4S & -- & 7.132 & no data  \\
 \hline
 1P & -- & 6.749 & no data   \\ 
 \hline
 2P & -- & 7.009 & no data   \\
 \hline
 1D & -- & 6.989 & no data  \\
 \hline\hline
\end{tabular}
\end{center}

\subsubsection*{Typical distance of the Cornell potential}

Up to now we have been working with natural units. Finding the correspondent value of the obtained parameters $C$ and $B$ in SI units, we acquire the typical distance $r_Q$, too.

\begin{center}
\begin{tabular}{||c |c|c |c|c||} 
 \hline\hline
  quarkonium &  ${\mu\ [\text{GeV\, c}^{-2}]}$ & ${B\ [\text{GeV\, fm}^{-1}]}$ & ${C\ [\text{GeV\, fm}]}$ & ${r_Q\ [10^{-16}\,\text{m}]}$ \\ [1ex]
  \hline
 $\text{c}\bar{\text{c}}$ & 0.64 & 1.633 & 0.175 & 3.28  \\ 
 \hline
 $\text{b}\bar{\text{b}}$ & 2.09 & 6.425 & 0.068 & 1.03 \\
 \hline
  $\text{c}\bar{\text{b}}$ & 0.97 & 3.067  & 0.119 & 1.97 \\
 \hline\hline
\end{tabular}
\end{center}

We have obtained $r_Q$ at the order of $10^{-16}\, \text{m}$, which is known to be roughly the size of the quark bound states \cite{size}. This further confirms the consistency of the results we have obtained.

\subsubsection*{Validity of the Pekeris-type approximation}

After fixing the parameters $B$ and $C$, we could proceed to the main goal of the present work: calculation of the non-commutative corrections to the masses. However, we should address the validity of the Pekeris-type approximation we have made. Let us remark that very few other works do so.

In section \eqref{sec2sub1}, we have claimed that the Pekeris-type approximation is valid only for the dimensionless radii ${y\approx1}$ in \eqref{WKBcorny}. We have found the numerical values of the variables $B$ and $C$, so we are ready to find numerically the limits of the integration in \eqref{WKBcorny}: $y_1$ and $y_2$. Our goal is to show that ${y_1\approx1\approx y_2}$, so the presumption ${y\approx1}$ is really realized. For synoptic reasons we are dealing with dimensionless quantities.

\begin{center}
\begin{tabular}{||c |c|c |c||} 
 \hline\hline
 $\text{c}\bar{\text{c}}$ meson &   &  ${b=1.883}$ & ${c=1.883}$\\ [1ex]
 \hline\hline
  state & $\epsilon\equiv\lr{\epsilon_{nl}}_{\text{WKB, PEK}}$ &  $y_{1}$  & $y_{2}$\\ [1ex]
 \hline\hline
 1S & 1.959 & 0.588 & 2.943  \\ 
 \hline
 2S & 4.030 & 0.230 & 3.301 \\
 \hline
 3S & 4.744 & 0.124 & 3.407 \\
 \hline
 4S & 5.071 & 0.105 &  3.453 \\
 \hline
 1P & 3.438 & 0.368 & 1.455  \\ 
 \hline
 2P & 4.512 & 0.166 & 1.656  \\
 \hline
 1D & 4.386 & 0.220 & 0.706  \\
 \hline\hline
\end{tabular}
\end{center}

Clearly, these values do stretch the condition ${y\approx 1}$ quite far. However, the values of masses they lead to are in a very good agreement with the experiment. As our goal is not in full understanding and control over our approximations, but we want find the NC corrections, we proceed as follows.

Consider \eqref{viete} as a starting point of a model to determine the masses of quarkonium states. Even though this came from WKB and Pekeris-type approximations, we can use it as a starting point for a model whose ultimate test is the agreement (or lack thereof) with the experiment. The model \eqref{viete} passes this test and we will use it in the following section. We leave the question of when and why this model works for the future.

\subsection{Charmonium, bottomonium and bottom-charmed meson masses in NC space}

The correction to the energy spectrum $\lr{\epsilon^{(2)}_{nl}}_{\text{WKB, PEK}}$ from the formula \eqref{emod} modifies the mass of the states \eqref{vsetko}. Taking this modification into account, we get the mass spectrum comprehending the non-commutativity of the space:  
\begin{align}
   M_{nl}^{\sigma}\,=& \,M_{nl}+\sigma^2 M_{nl}^{(2)}+\dots=\nonumber \\
   \,=& \,\lr{(m_1+m_2)-\frac{2\mu}{\hbar^2}\slr{\frac{2C}{n+\frac12+\sqrt{\frac{2\mu}{\hbar^2}C\sqrt{\frac{C}{B}}+\lr{l+\frac12}^2}}}^2+3\sqrt{BC}}+\nonumber \\
   &+\sigma^2\frac{\hbar^2}{2\mu}\frac{B}{C}\left(
   \frac{b \left(105 b^2+62 b c+9 c^2\right)+4b(c+3 b)l (l+1)}{8\left[n+\frac12+\sqrt{b+( l+\frac12)^2}\right]^2}
   -\frac{b}{4}\lr{15b+4c}+\right. \nonumber\\ 
&+\frac{b (c+3 b)^4}{8\sqrt{b+( l+\frac12)^2} \left[n+\frac12+\sqrt{b+( l+\frac12)^2}\right]^5}\left.-\frac{(45 b-c) (c+3 b)^3}{64 \left[n+\frac12+\sqrt{b+( l+\frac12)^2}\right]^4}\vphantom{\frac{1}{2}}\right)+\dots \ ,\label{}
\end{align}
where we remind the reader that ${b=2\mu Br_Q^3/\hbar^2}$, ${c=2\mu Cr_Q/\hbar^2}$.

We calculate this modification of the mass spectrum $\sigma^2 M_{nl}^{(2)}$ for all the examined states of all the three systems. Since the potential effect of the non-commutativity of the space is on the $39^{\text{th}}$ decimal place in the mass spectrum of the mesons, this modification is beyond the limit of accuracy of any current measurement. For this reason we used the non-perturbed mass spectra to find the parameters $B$ and $C$, and with the help of these parameters give an estimation for the leading correction in the non-commutative case.

We obtain corrections for the charmonium meson given by the following table.

\begin{center}
\begin{tabular}{||c |c||} 
 \hline\hline
 NC $\text{c}\bar{\text{c}}$ meson &  ${b=c=1.883}\ ,\ \ {\sigma^2\approx0.93\, \times\, 10^{-39}} $\\ [1ex]
 \hline\hline
  state & correction to the mass spectrum $\sigma^2 M_{nl}^{(2)}$ [GeV]\\ [1ex]
 \hline\hline
 1S & ${\color{white}+}0.522\, \sigma^2$   \\ 
 \hline
 2S & $-1.422\, \sigma^2$ \\
 \hline
 3S & $-2.613\, \sigma^2$ \\
 \hline
 4S & $-3.301\, \sigma^2$ \\
 \hline
 1P & $-0.456\, \sigma^2$  \\ 
 \hline
 2P & $-1.936\, \sigma^2$  \\
 \hline
 1D & $-1.062\, \sigma^2$  \\
 \hline\hline
\end{tabular}
\end{center}

Note that we get a positive correction only for the 1S state. 

We obtain corrections for the bottomonium given by the following table.

\begin{center}
\begin{tabular}{||c |c||} 
 \hline\hline
 NC $\text{b}\bar{\text{b}}$ meson &  ${b=c=0.750}\ ,\ \ {\sigma^2\approx9.43\, \times\, 10^{-39}} $\\ [1ex]
 \hline\hline
  state & correction to the mass spectrum $\sigma^2 M_{nl}^{(2)}$ [GeV]\\ [1ex]
 \hline\hline
 1S & $-0.261\, \sigma^2$   \\ 
 \hline
 2S & $-1.289\, \sigma^2$ \\
 \hline
 3S & $-1.753\, \sigma^2$ \\
 \hline
 4S & $-1.973\, \sigma^2$ \\
 \hline
 1P & $-0.738\, \sigma^2$  \\ 
 \hline
 2P & $-1.480\, \sigma^2$  \\
 \hline
 1D & $-1.042\, \sigma^2$  \\
 \hline\hline
\end{tabular}
\end{center}

The setup of \cite{NC_bb,NC_cc} allowed for a direct use of the standard perturbation theory in quantum mechanics, without any need of WKB or Pekeris-type approximations. Both of these works obtained leading correction to the energy spectrum at the second order in non-commutative parameter, as did we, but with a different $n$ and $l$ dependence. Interestingly, for some values of the quantum numbers the correction has diverged, whereas our results are under control for all values. Let us remind the reader that the setup of that work uses a different version of the NC Hamiltonian than \eqref{NCschr}.

Finally, let us give the results for the $\text{c}\bar{\text{b}}$ meson.

\begin{center}
\begin{tabular}{||c |c||} 
 \hline\hline
 NC $\text{c}\bar{\text{b}}$ meson &  ${b=c=1.174}\ ,\ \ {\sigma^2\approx2.58\, \times\, 10^{-39}} $\\ [1ex]
 \hline\hline
  state & correction to the mass spectrum $\sigma^2 M_{nl}^{(2)}$ [GeV]\\ [1ex]
 \hline\hline
 1S & $-0.001\, \sigma^2$   \\ 
 \hline
 2S & $-1.442\, \sigma^2$ \\
 \hline
 3S & $-2.210\, \sigma^2$ \\
 \hline
 4S & $-2.609\, \sigma^2$ \\
 \hline
 1P & $-0.672\, \sigma^2$  \\ 
 \hline
 2P & $-1.777\, \sigma^2$  \\
 \hline
 1D & $-1.147\, \sigma^2$  \\
 \hline\hline
\end{tabular}
\end{center}

\subsubsection*{Maximum possible value of the of the non-commutative parameter $\lambda$}

Although, we have claimed that $\lambda$ is expected to be at the Planck scale, i.e. $10^{-35}\,\text{m}$, it is just a hypothesis which cannot be obtained from experimental data. We therefore take the uncertainty of the measurement into account: the uncertainty of the mass $M_{nl}$ can be an upper bound of the non-commutative mass spectrum correction $\sigma^2 M_{nl}^{(2)}$. From this equality we acquire the maximum possible value of $\lambda$.

Among all the mentioned particles in this paper the mass of the $J/\psi$(1S) particle -- a $\text{c}\bar{\text{c}}$ meson -- is measured the most precisely: ${M_{00}=3096.900\pm0.006\, \text{MeV}}$, see  \cite{pdg2}. So, we use its uncertainty to determine the upper value of $\lambda$. With the help of the correction to the mass spectrum $\sigma^2 M_{00}^{(2)}$ for the $\text{c}\bar{\text{c}}$ meson we get the following inequality
\begin{equation}
    0.522\, \lr{\lambda^2\frac{B}{C}}\, \text{GeV}\leq 0.006\, \text{MeV}
\end{equation}
Expressed with natural units we get
\begin{equation*}
    \lambda\leq 5.64\, \times\, 10^{-3}\, \text{GeV}^{-1}\ ,
\end{equation*}
which corresponds to the value in SI units
\begin{equation}\label{limit}
    \lambda\leq 1.11\, \times\, 10^{-18}\, \text{m}\ .
\end{equation}
Both \cite{NC_bb} and \cite{NC_cc} have considered the hyper-fine splitting of the quarkonium states to determine the upper bound on the characteristic distance of the non-commutativity, obtaining a value greater by two orders of magnitude than \eqref{limit}. Since the experimental error in the measurement of $J/\psi$(1S) is much smaller than the splitting, we obtain a stricter bound on $\lambda$. However, even the bound we obtain is still a very mild one.

\color{black}


\section{Conclusions}\label{sec5}

We have studied the consequence of non-commutativity of space on the masses of bound states of heavy quarks. Working under the non-relativistic assumption and using the WKB and slightly modified Pekeris-type approximations, we have derived the masses of the considered mesons (as the commutative ${0^{\mathrm{th}}}$ order term) and the first non-trivial correction due to the non-commutative structure. Our model had two free parameters, which have been fixed using the experimentally observed masses and gave the rest of the masses in a good agreement with the experiment. The leading NC correction is proportional to $(\lambda/r_Q)^2$, where $\lambda$ is the characteristic distance of the space non-commutativity and $r_Q$ is the characteristic distance of the quarkonium state.

For $\lambda$ at the order of Planck length, where the quantum structure of space-time is expected to be significant from the quantum gravity considerations, we obtained relative correction of the mass at the order of $10^{-39}$. Considering the most precisely measured mass, we have obtained the upper bound on $\lambda$ at the order of $10^{-18}\,\textrm{m}$.

Our hope was that the quarkonium states will turn out to be a (relatively) good probe for the quantum structure of spacetime. Since the typical radius of the states $r_Q$ is much smaller than the Bohr radius, we do get a significantly larger shift than in the case of hydrogen atom \cite{svk}, but we are still far from any reasonably measurable contribution. The main issue is that the relative corrections to the masses turned out to be of the second order in $\lambda/r_Q$. So, in the future, it would be interesting to look further for a system where the correction is of the first order.


\acknowledgments
We would like to thank Samuel Kov\'a\v cik, Peter Mat\'ak, Boris Tom\'a\v sik, Zuzana Ku\v cerov\'a and Alexander Rothkopf for helpful discussions.

This work was supported by the \emph{Alumni FMFI} foundation as a part of the \emph{N\'{a}vrat teoretikov} project and by VEGA 1/0703/20 grant \emph{Quantum structure of spacetime}.

\end{document}